\documentclass[aps,prd,reprint,preprintnumbers,superscriptaddress,showpacs,twocolumn]{revtex4-1}
\usepackage{latexsym,graphicx,amssymb,amsmath,mathrsfs}
\usepackage{setspace,bm}
\usepackage[breaklinks, colorlinks=true, pdfstartview=FitV, linkcolor=red, citecolor=blue, urlcolor=blue]{hyperref}
\usepackage[usenames]{color}
\usepackage{latexsym}
\usepackage{epstopdf}
\usepackage{mathtools}
\usepackage{amssymb}

\usepackage[utf8]{inputenc}

\usepackage{tikz}
\usetikzlibrary{shapes,arrows}

\allowdisplaybreaks[1]
\usepackage[normalem]{ulem}  

\renewcommand\sout{\bgroup \color{red} \ULdepth=-.5ex \ULset}

\begin{document}
\preprint{KUNS-2822}

\title{
Path optimization for $U(1)$ gauge theory with complexified parameters
}

\author{Kouji Kashiwa}
\email[]{kashiwa@fit.ac.jp}
\affiliation{Fukuoka Institute of Technology, Wajiro, Fukuoka 811-0295,
Japan}

\author{Yuto Mori}
\email[]{mori@ruby.scphys.kyoto-u.ac.jp}
\affiliation{Department of Physics, Faculty of Science, Kyoto
University, Kyoto 606-8502, Japan}

\begin{abstract}
 In this article, we apply the path optimization method to handle the complexified parameters in the 1+1 dimensional pure $U(1)$ gauge theory on the lattice.
 Complexified parameters make it possible to explore the Lee-Yang zeros which helps us to understand the phase structure and thus we consider the complex coupling constant with the path optimization method in the theory.
 We clarify the gauge fixing issue in the path optimization method; the gauge fixing helps to optimize the integration path effectively.
 With the gauge fixing, the path optimization method can treat the complex parameter and control the sign problem.
 It is the first step to directly tackle the Lee-Yang zero analysis of the gauge theory by using the path optimization method.
\end{abstract}

\maketitle


\section{Introduction}

Exploring the phase structure of theories and models with finite external parameters such as the temperature ($T$), the chemical potential ($\mu$) and the external magnetic field ($B$) is an important subject to understand our universe. For example, the phase structure of quantum chromodynamics (QCD) at finite $T$, $\mu$ and $B$ is directly related to the early universe, current heavy ion collision experiments, neutron star physics and so on; see Ref.~\cite{Fukushima:2010bq} as an example.

One of the interesting approaches to investigate the phase structure is the Lee-Yang zero analysis~\cite{Lee:1952ig,biskup2000general}. In the analysis, we complexify external parameters and search zeros of the partition function in the complex plane of the external parameters. Then, an approaching tendency of zeros to the real axis indicates the existence of the phase transition because singularities of the partition function are the origin of the ordinary phase transition.
Particularly, the experimental observation~\cite{peng2015experimental} and the quantum computation by using a quantum computer~\cite{PhysRevA.100.022125} for the zeros are currently possible and thus the analysis has attracted much more attention recently.

There have been some attempts to perform the Lee-Yang zero analysis in the gauge theory; an interesting example is QCD with the complexified $\mu$~\cite{Nakamura:2013ska,Nagata:2014fra,Wakayama:2018wkc}. In the calculation, one first gathers numerical data at finite imaginary chemical potential ($\mu_\mathrm{I}$) and after they construct the grand canonical partition function with the complex $\mu$ by using the Fourier transformation and the fugacity expansion; see Ref.\,\cite{Roberge:1986mm}.
However, the Fourier transformation becomes much more difficult with decreasing $T$ because the oscillation becomes severer and thus this approach cannot tell us the phase structure at low $T$; for example, see Ref.\,\cite{Fukuda:2015mva}. The reason why we use the imaginary chemical potential to perform the Lee-Yang zero analysis in QCD is that there is the sign problem at complexified external parameters and then the Monte-Carlo calculation sometimes fails; see Ref.\,\cite{deForcrand:2010ys} for details of the sign problem and  Ref.\,\cite{Kashiwa:2019ihm} for details of the imaginary chemical potential. If we can directly perform the Monte-Carlo calculation with finite complexified parameters, there is the possibility that we can better understand properties of the phase structure via the Lee-Yang zero analysis.
In addition, such complexification of chemical potential may be related to the investigation of the confinement-deconfinement transition at finite density~\cite{Kashiwa:2015tna,Kashiwa:2017yvy}.

In the Lefschetz thimble and path optimization methods, dynamical variables are complexified and then the integral path and/or the configurations are generated such that those obeying the sign-problem weaken manifold.
Since these approaches can weaken the sign problem and thus it is natural to expect that these approaches can be applied to explore the system with complexified external parameters. In this study, we concentrate on the application of the path optimization method to the system with complexified parameters.

The path optimization method is based on the standard path integral formulation with the complexification of dynamical variables~\cite{Mori:2017zyl,Mori:2017pne};
the actual procedure is performed as follows:
\begin{enumerate}
  \item The cost function, which reflects the seriousness of the sign problem, is prepared.
  \item Dynamical variables are Complexified.
  \item The integral path in the complex domain is modified to minimize the cost function.
\end{enumerate}
After taking the prescription, we can have a better integral path (manifold) which has larger $|e^{i\theta}|$; $0 \le |e^{i\theta}| \le 1$ is the average phase factor which is responsible for the seriousness of the sign problem.
Thanks to Cauchy's integral theorem, the modified integral path provides us the same result as that obtained on the original integral path if there are no poles or cuts between the modified and original paths and the infinite regions of the integral path do not contribute to the results.
There are some attempts to apply the method to various quantum field theories and models, e.g., the complex $\lambda \phi^4$ theory~\cite{Mori:2017pne},
the Polyakov-loop extended Nambu--Jona-Lasinio model~\cite{Kashiwa:2018vxr,Kashiwa:2019lkv} and the $0+1$ dimensional QCD~\cite{Mori:2019tux}.

In this article, we apply the path optimization method to deal with the complexified parameters.
We here employ one plaquette system in the 1+1 dimensional $U(1)$ gauge theory with complexified coupling constant on the lattice; some results for this theory are obtained by using a modification of the integral path, see Refs.\,\cite{Pawlowski:2020kok,Detmold:2020ncp}.
In Sec.\,\ref{Sec:formulation}, we show the formulation of the theory on the lattice and the explanation of the path optimization method.
Numerical results are shown in Sec.\,\ref{Sec:numerical}.
Section\,\ref{Sec:summary} is devoted to summary.


\section{Formulation}
\label{Sec:formulation}

In this section, we summarize detailed formulation of the 1+1 dimensional $U(1)$ lattice gauge theory and explain how we apply the path optimization method to the theory.
We here consider the one plaquette system and thus the following formulation is corresponding to the system.

\subsection{Action and partition function}
The Wilson's plaquette action~\cite{Wilson:1974sk} for only one plaquette in the case of the $U(1)$ gauge theory is written as
\begin{align}
    S_\mathrm{G} &= \frac{\beta }{2} \Bigl\{ P + P^{-1} \Bigr\},
    \label{Eq:action}
\end{align}
where
$\beta = 1/g^2$ is the lattice gauge coupling constant and $P$ ($P^{-1}$) denotes the plaquette (its inverse). The definition of $P$ is given by
\begin{align}
    P := U_1 \, U_2 \, U^{-1}_3 \, U^{-1}_4,
\end{align}
where $U_n$ are the $U(1)$ link variables defined as
\begin{align}
    U_n &:= e^{ig A_n} \in \mathrm{U(1)},
\end{align}
here $A_n$ denotes the $U(1)$ gauge field, $g^2 = 1/\beta$ and $n = 1,2,3,4$.

The present theory can be analytically solved as
\begin{align}
    {\cal Z} &= \int \prod_{n} dU_n \, e^{-S_\mathrm{G}}
    = I_0(\beta),
    \label{Eq:analytic}
\end{align}
where $I_0$ denotes the modified Bessel function of the first kind.
It should be noted that we can obtain analytic result of the partition function for any system volume with periodic or open boundary conditions on the lattice~\cite{Balian:1974ts,Pawlowski:2020kok,Detmold:2020ncp}.
For real $\beta$, $I_0(\beta)$ is always positive and thus there are no zeros of $\mathcal{Z}$ in Eq.\,(\ref{Eq:analytic}), but the gauge coupling constant is now complex, $\beta \in \mathbb{C}$, and thus the partition function can be $0$ which is nothing but the Lee-Yang zeros.
These zeros play an important role to understand the phase structure.

For the gauge theory, the action is invariant under the gauge transformation.
In this case, to use the gauge transformation, one can reduce the degree of freedom to $n_\mathrm{deg} = 1, \cdots, 4$;
\begin{align}
U_n =
\begin{cases}
U_n & n=1,\cdots,n_{\mathrm{deg}}\\
{\mathbf I}   & \mathrm{otherwise}.
\end{cases}
\end{align}

\subsection{Path optimization method}
In the path optimization method, we extend dynamical variables from real ($t \in \mathbb{R}$) to complex ($z \in \mathbb{C}$).
In the present case, we need to extend the plaquette and the link variable as
\begin{align}
    {\cal P} &= {\cal U}_1 \, {\cal U}_2 \, {\cal U}_3^{-1} \, {\cal U}_4^{-1},
\\
    {\cal U}_n &= e^{ig {\cal A}_n} =: U_n \, e^{z_n},
\end{align}
where ${\cal A}_n \in \mathbb{C}$ and then  $z_n \in \mathbb{R}$ represents the modification of the integral path.
To represent $z_n$, we employ the artificial neural network as the model to generate the integral path.
We here use the simple neural network which has the mono input, hidden and output layers. In this network, the variables in the hidden layer nodes ($y_j$) and the output variables ($z_n$) are given as follows.
\begin{align}
    & z_n  = \omega_n F(w^{(2)}_{jn} y_j + b_j ),
    \nonumber\\
    & y_j = F(w^{(1)}_{ij} t_i + b_j), \label{Eq:FNN}
\end{align}
where $t_i$ denotes the parametric variable which is set to $\mathrm{Re}~U_{n'}$, $\mathrm{Im}~U_{n'}$,
$i=1, \cdots, 2 \times n_\mathrm{deg}$, $w$, $b$ and $\omega$ are parameters of the neural network (weight and bias) and
$F$ is so called the activation function. In this study, we chose the tangent hyperbolic function as the activation function.

To perform the path optimization, we need the cost function (${\cal F}$); we here use
\begin{align}
    {\cal F}[z(t)] &= \int d^n t \, |e^{i\theta(t)}-e^{i\theta_0}|^2 \times |J(t) \, e^{-S_\mathrm{G}(t)}|, 
\end{align}
where $J(t)$ is the Jacobian, $\theta_0$ denotes the phase of the average phase factor and $\theta(t)$ is the phase of $J(t) e^{-S_\mathrm{G}(t)}=e^{i\theta(t)} |J(t) e^{-S_\mathrm{G}(t)}|$. 
Of course, we do not know the actual value of $\theta_0$ except with $\beta \in \mathbb{R}$, $\theta_0=0$, and thus we replace $\theta_0$ with $\langle \theta_\mathrm{pre} \rangle_\mathrm{EMA}$ where $\langle \theta_\mathrm{pre} \rangle_\mathrm{EMA}$ is the exponential moving average (EMA) of the phase obtained in the previous optimization steps.
Minimization of the above cost function makes phases of $e^{-S_\mathrm{G}}$ as a function of $z$ take similar values on the modified integral path when the regions are relevant to the final result. In other words, there is no need to care for the phase of the Boltzmann weight in irrelevant regions which should be automatically suppressed.

Since there is the sign problem in the case of the complex coupling constant by definition, we use the phase reweighting to perform the Monte-Carlo calculation as
\begin{align}
    \langle {\cal O} \rangle
    &= \frac{\langle {\cal O} e^{i\theta} \rangle_\mathrm{pq}}
            {\langle e^{i\theta} \rangle_\mathrm{pq}},
\label{Eq:pq}
\end{align}
where ${\cal O}$ represents any operator and $\langle \cdots \rangle_\mathrm{pq}$ means the phase quenched expectation values where $|Je^{-S_\mathrm{G}}|$ is used as the Boltzmann weight. Since the Boltzmann weight, $|Je^{-S_\mathrm{G}}|$, is now real, we can perform the Monte-Carlo calculation exactly.
It should be noted that we are not restricted to the original integral path in the estimation of Eq.\,(\ref{Eq:pq}) unlike the ordinary reweighting calculation.

The machine learning technique was first introduced to the path optimization method in Ref.~\cite{Mori:2017pne} to represent the modified integral path with a weaker sign problem. The machine learning technique was also introduced  to the generalized Lefschetz thimble method~\cite{Alexandru:2015sua} to learn the integral manifold where the sign problem is mild in Ref.\,\cite{Alexandru:2017czx} a few days before Ref.~\cite{Mori:2017pne}.

\subsection{Setting of numerical calculation}
Numbers of the unit in the input, hidden and output layers are $N_\mathrm{input} = 2 \times n_\mathrm{deg}$, $N_\mathrm{hidden}=10$ and $N_\mathrm{output}=n_\mathrm{deg}$, respectively.
To determine the parameters in the neural network, we optimize these by using the ADADELTA~\cite{zeiler2012adadelta}, one of the stochastic gradient methods, as an optimizer with the Xavier initialization~\cite{glorot2010understanding}, the batch normalization~\cite{ioffe2015batch}, the mini-batch training and the exponential moving average; see Ref.\,\cite{Mori:2017pne} for details of the optimization.

Actual configurations are generated by using the path optimization method with the hybrid Monte-Carlo (HMC) method~\cite{Duane:1987de} in the systems which includes the single plaquette with the open boundary condition.
It should be noted that we here use the HMC with the replica exchange method (exchange HMC)~\cite{swendsen1986replica,Fukuma:2017fjq} because there is the global sign problem even on the modified integral path~\cite{Fujii:2013sra} which means that there are some separated regions on the modified integral path relevant to the integration. Integration over these separated regions is quite difficult to pick up by using ordinary HMC: 
We prepare the $N_{rep}$ replicas characterized by the parameters in neural network as
\begin{align}
  C_x = \frac{x}{N_{rep}}C,
\end{align}
where $x$ means the replica number, $x=1,\cdots,N_{rep}$, and $C$ represents the parameters of the optimized neural network ($C=w, b, \omega$ in Eq.~(\ref{Eq:FNN})).
We set $N_{rep} = 5$ in the numerical calculation.
We use the exchange probability of the replicas $P(U_x \leftrightarrow U_{x'})$ as
\begin{align}
  P(U_x \leftrightarrow U_{x'}) &= \min \left( 1, \frac{P(U_{x};C_{x'})P(U_{x'};C_x)}{P(U_{x};C_x)P(U_{x'};C_{x'})} \right),\nonumber \\
  P(U_x; C_x) &= |J(U_x;C_x) e^{-S_G(\mathcal{U}(U_x;C_x))}|.
\end{align}
In Ref.\,\cite{Mori:2017pne}, we used the sampling of configurations based on the symmetry of the modified integral path, but the method is only available if we know the good modified integral path which can have the symmetry.
In this paper, we employ the exchange HMC method to generally perform the path optimization.

The expectation values are calculated with $2500$ configurations and then the corresponding errors are estimated by using the Jackknife method with the bin-size $50$.
For the parameter of the theory, we set $\beta = \beta_\mathrm{R} + i \beta_\mathrm{I}$ with $\beta_\mathrm{R}, \, \beta_\mathrm{I} \in \mathbb{R}$.

\begin{figure}
\centering
\includegraphics{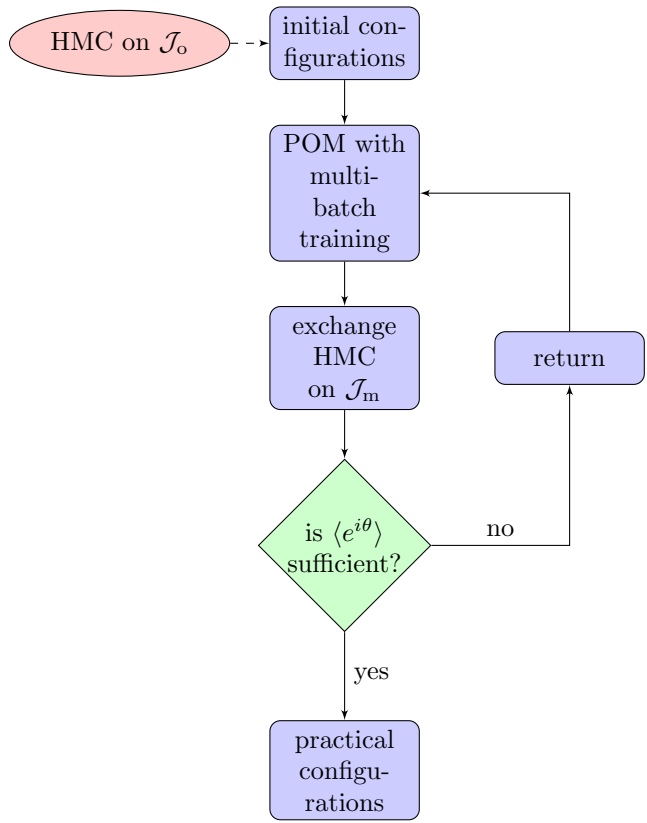}
\caption{The flowchart of the algorithm to generate practical configurations in this work. Symbols, ${\cal J}_\mathrm{o}$ and ${\cal J}_\mathrm{m}$, denote the original and the modified integral path, respectively. The closed loop in the flowchart is the one cycle of the optimization procedure of the integral path.}
\label{Fig:fc}
\end{figure}

Figure~\ref{Fig:fc} shows the flowchart of the algorithm to generate practical configurations where ${\cal J}_\mathrm{o}$ and ${\cal J}_\mathrm{m}$ are the original and modified integral path, respectively.

\section{Numerical results}
\label{Sec:numerical}

In this section, we show the numerical results of the 1+1 dimensional $U(1)$ gauge theory on the lattice with the complex coupling by using the path optimization method.
Here, we show the results of the $1+1$ dimensional $U(1)$ gauge theory only with single plaquette; i.e. the simplest setting of the theory.
Actually, it is nothing but the $n_\mathrm{deg}$-dimensional integral.

\begin{figure}[t]
 \centering
 \includegraphics[width=0.43\textwidth]{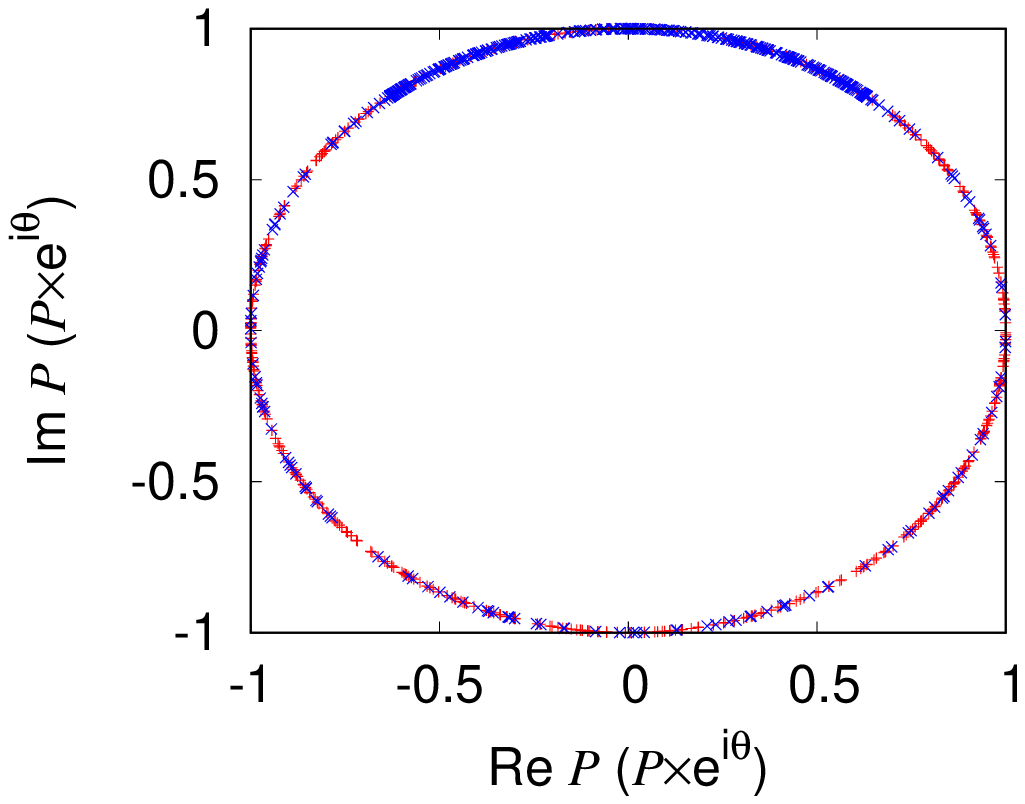}\\
 ~~~~~\includegraphics[width=0.4\textwidth]{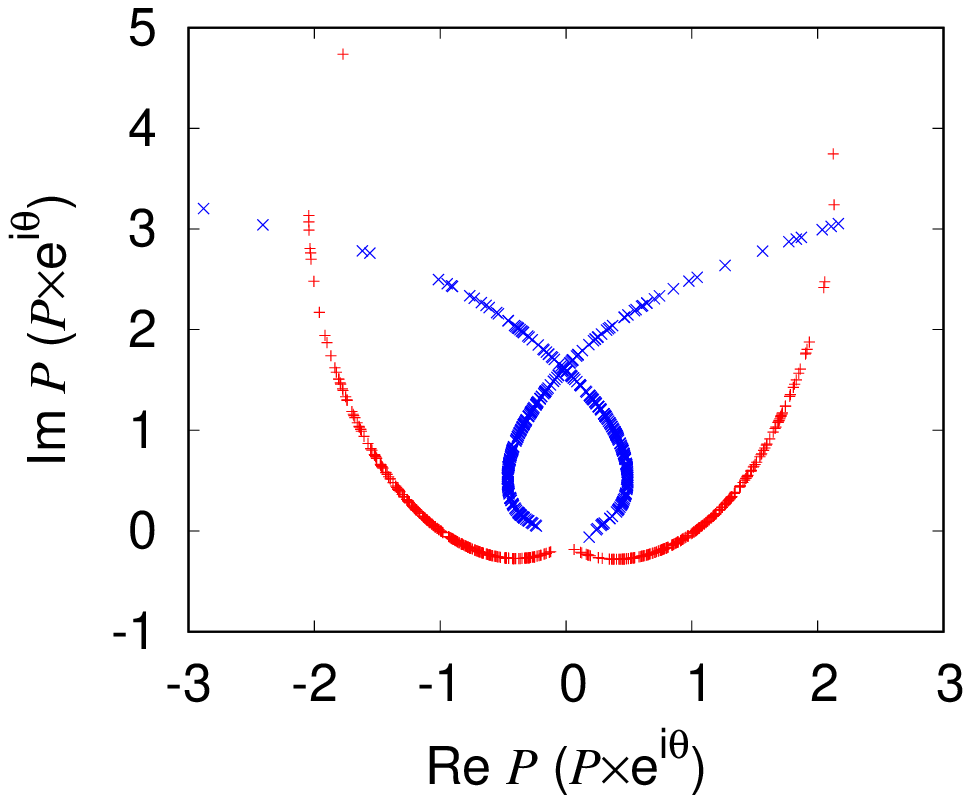}
 \caption{
The scatter plot of ${\cal P}$ and ${\cal P}\times e^{i\theta}$ on the $X$-$Y$ plane with $\beta = 2i$ where $X=\mathrm{Re}\,{\cal P}$ and $Y=\mathrm{Im}\,{\cal P}$.
 The top and bottom panels show the results without and with the path optimization.
 Plus signs and crosses indicate ${\cal P}$ and ${\cal P}\times e^{i\theta}$, respectively.
 }
\label{Fig:sp}
\end{figure}

\begin{figure}[t]
 \centering
 \includegraphics[width=0.43\textwidth]{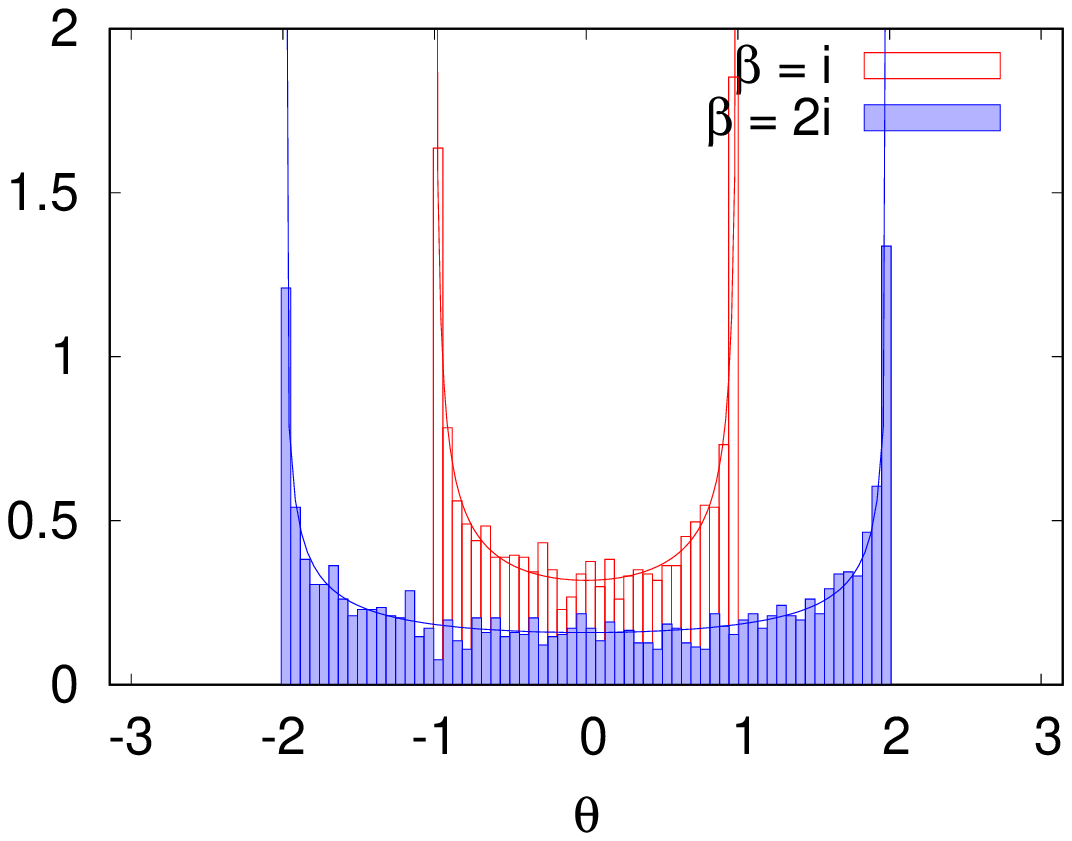}\\
 ~~~\includegraphics[width=0.41\textwidth]{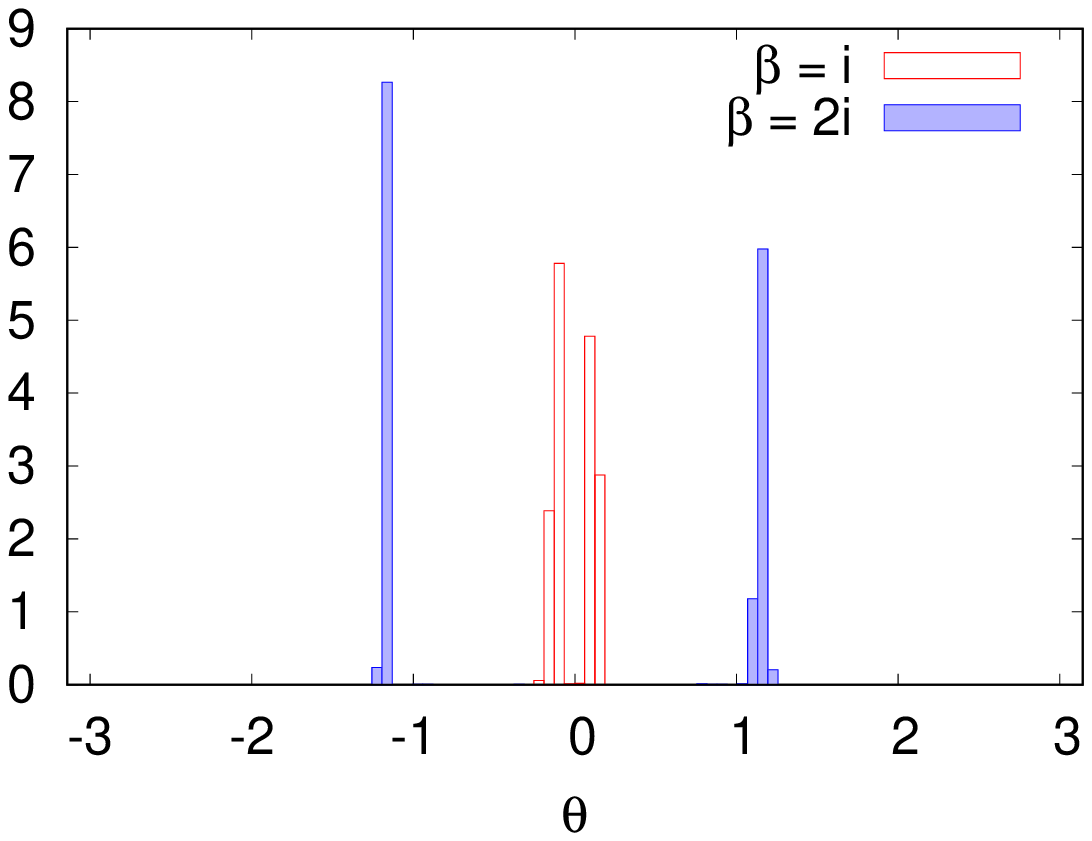}
 \caption{The normalized histogram of the phase, $\theta$, for $\beta = i$ and $2i$.
 The top and bottom panels show the results without and with the path optimization.
 The line in the top panel shows the probability distribution on the original path, $P(\theta) = [ \pi \beta_\mathrm{I} \sin\{\arccos(\theta/\beta_I)\}]^{-1}$.
 }
\label{Fig:hist}
\end{figure}

Figure \ref{Fig:sp} shows the scatter plot on the $\mathrm{Re} \, {\cal P}$-$\mathrm{Im} \, {\cal P}$ plane with $\beta = 2i$.
Here, we show the results with the gauge fixing:
\begin{align}
U_n &=
    \begin{cases}
    U_n & ~~n=1\\
    {\mathbf I} & ~~n \neq 1
  \end{cases}
  .
\end{align}
We can clearly see the modification of the integral path from fig.\,\ref{Fig:sp}.
In addition, we can see the bias of the distribution in
$\mathcal{P}\times e^{i \theta}$ which plays a crucial role in the calculation of
$\langle {\cal P} \rangle$; this bias makes
$\langle {\cal P} \rangle$ finite because the phase quenched expectation value of ${\cal P}$ becomes $0$.
The histogram of the phase for the case of Fig.\,\ref{Fig:sp} is shown in Fig.\,\ref{Fig:hist}.
On the original integral path, $\theta$ distribution is widely spread, but the distribution on the modified integral path is well localized; we can generate configurations strongly localized around two separated regions.
The replica exchange method well works in both cases.

\begin{figure}[t]
 \centering
 \includegraphics[width=0.52\textwidth]{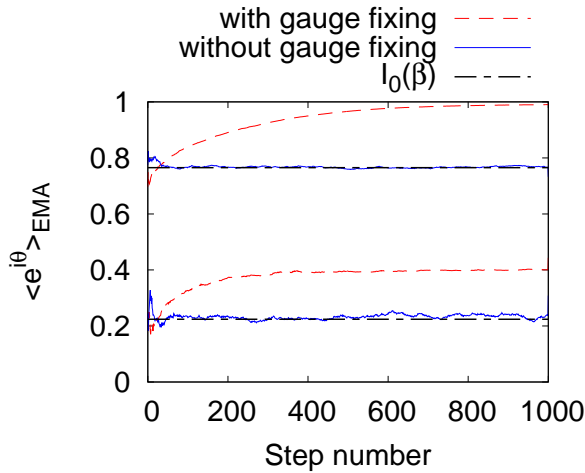}\\
 \caption{The average phase factor, $\langle e^{i\theta} \rangle_\mathrm{EMA}$, during the optimization with and without the gauge fixing in one epoch. The dash-dot line shows the exact average phase factor on the original path.
  In the case with the gauge fixing, we here perfectly fix the gauge degree of freedom; i.e. $n_\mathrm{deg}=1$.
 Upper- and bottom-side lines are results with $\beta =i$ and $2i$, respectively. In the case with $\beta=2i$, the serious global sign problem exists.
 }
\label{Fig:af-t}
\end{figure}

\begin{figure}[t]
 \centering
 \includegraphics[width=0.45\textwidth]{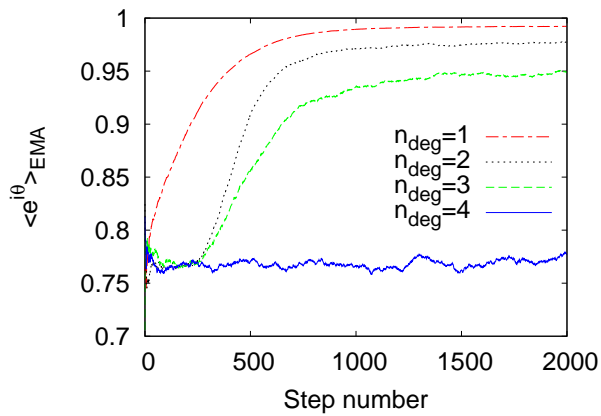}\\
 \caption{The average phase factor, $\langle e^{i\theta} \rangle_\mathrm{EMA}$, during the optimization with and without the gauge fixing in two epochs with $\beta =i$.}
\label{Fig:com}
\end{figure}

Figure\,\ref{Fig:af-t} shows the average phase factor, $\langle e^{i\theta} \rangle_\mathrm{EMA}$, as a function of the optimization step in one epoch with $\beta = i$ and $2i$; one epoch is defined so that one sequence of the mini-batch training is finished.
Here, we estimate the average phase factor by using EMA.
From the figure, we can clearly see that the average phase factor cannot be enhanced without the gauge fixing. 
With the gauge fixing, the average phase factor approaches $1$ with $\beta =i$.
In the case with $\beta=2i$, there is the serious global sign problem as shown in Fig.\,\ref{Fig:hist} and thus we have the upper limit of the improvement, but we can well enhance the average phase factor via the path optimization.
It should be noted that the modified integral path sometimes provides the expectation value with the larger error-bar compared with the original one even if the average phase factor is enhanced.
This indicates that there is the competition between the improvement of the original sign problem and the modification of the integral path which is responsible to the statistical error via the path optimization method.
In addition, we can see from Fig.\,\ref{Fig:com} that the average phase factor becomes larger with reduction of $n_\mathrm{deg}$ by using the gauge fixing;
it may be related to the fact that we have larger degree of freedom without the gauge fixing and then the neural network cannot show sufficient performance to optimize the suitable integral path.

\begin{figure}[h]
 \centering
 \includegraphics[width=0.45\textwidth]{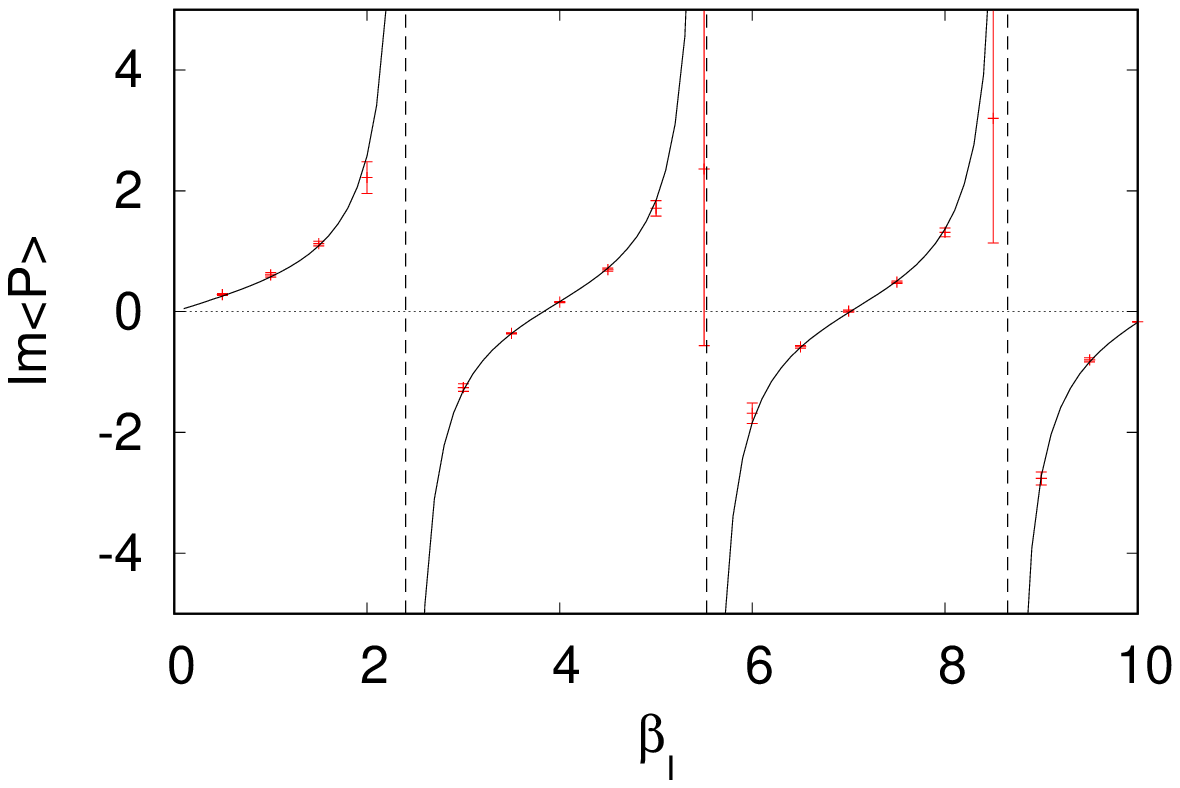}\\
 \includegraphics[width=0.45\textwidth]{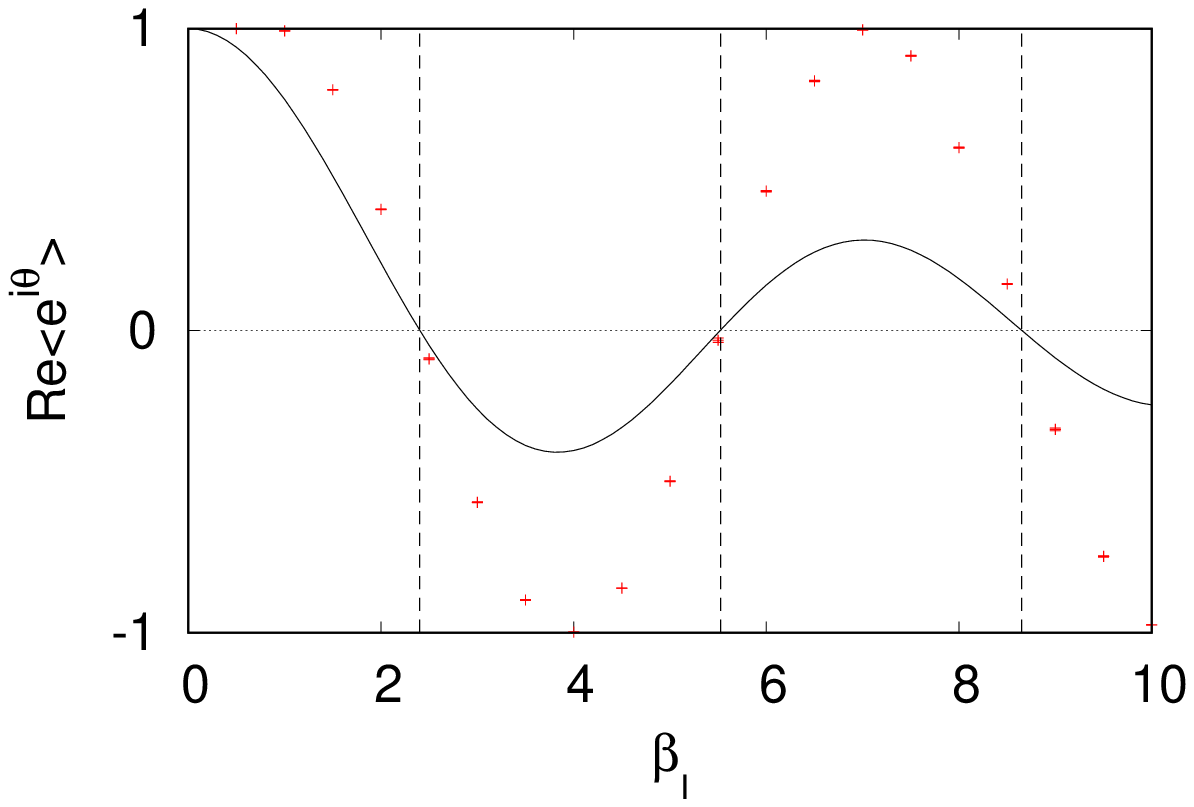}\\
 \caption{The top and bottom panels show the $\beta_\mathrm{I}$-dependence of the expectation value of the plaquette and the average phase factor, respectively.
 Symbols show numerical results obtained via the path optimization with the gauge fixing and solid lines denote the analytic results;
 the solid line in the bottom panel is corresponding to the results on the original integral path.}
\label{Fig:ar}
\end{figure}

For the reader's convenience, we finally show the expectation value of the plaquette as a function of $\beta_\mathrm{I}$ with $\beta_\mathrm{R}=0$ where zeros exist in Fig.\,\ref{Fig:ar}.
It is clearly seen that the modified integral path reproduces the analytic result except the region near the partition function zeros. At $\beta_\mathrm{I}\sim 2.4, 5.5, 8.6$, we have the zeros and then there should be the strong modification of the integral path and/or the serious global sign problem.


\section{Summary}
\label{Sec:summary}

In this study, we have considered the 1+1 dimensional $U(1)$ lattice gauge theory which has the single plaquette with the complex coupling constant as a first step to directly investigate the Lee-Yang zeros in the gauge theory. We have estimated how the average phase factor is improved via the modification of integral variables.
Since there is the sign problem when the coupling constant is complex, we employ the path optimization method to perform the Monte-Carlo calculation.
To represent the modified integral path in the complex domain of the integral variables, we employ the artificial neural network.

We have shown that the modification of the integral path represented by using the neural network can well enhance the average factor if we impose the gauge fixing to the theory, but we cannot without the gauge fixing; this suggests the importance of the gauge fixing to control the sign problem in the path optimization method on the lattice.
This issue is demonstrated in the system of the single plaquette.
Also, we have checked that the replica exchange method well works to generate configuration localized in the well separated regions which are realized via the path optimization.
From these results, we have clarified how to use the path optimization method in the gauge theory. It should be important in the application of the path optimization method to the more complicated gauge theory such as QCD.

In the present study, we have restricted the system size to be small because we are interested in the possibility of applying the path optimization method to the system with complexified external parameters.
Usually, the sign problem becomes exponentially worse in terms of the system volume; there is the competition between the exponential suppression of it from the system volume and the enhancement of it from the optimization.
In the larger volume case, we should introduce some methods to reduce the numerical cost to calculate the Jacobian, whose numerical cost is proportional to the square of the system volume.
Examples of promising methods are the diagonal ansatz of the Jacobian \cite{Alexandru:2018fqp} and the nearest-neighbor lattice-sites ansatz \cite{Bursa:2018ykf}.
We will revisit this issue in our future work.

This study is a first step in the path optimization method to explore the phase structure of the gauge theory in the complexified parameter space which is important to understand properties of the phase transition; e.g. for investigation of the distribution of the Lee-Yang zeros.
In the present work, we employ the simple gauge theory, but we believe that it sheds light on the complexification of the integral variables and also the parameters.
In the future, we will apply the path optimization method to the $SU(2)$ gauge theory with the complex coupling constant.
It was reported that the complex Langevin method fails in some parameter regions; see Ref.\,\cite{Makino:2015ooa}.

\begin{acknowledgments}
We are very grateful to A. Ohnishi and Y. Namekawa for their careful reading of the manuscript and useful comments.
We appreciate S. Hawkins for reading the manuscript via the English writing correction service provided from Fukuoka Institute of Technology.
We acknowledge the Lattice Tool Kit (LTKf90) since our code is in part based on LTKf90~\cite{Choe:2002pu}.
This work is supported by the Grants-in-Aid for Scientific Research from JSPS (No. 18K03618, 18J21251 and 19H01898).
\end{acknowledgments}

\bibliography{ref.bib}

\begin{thebibliography}{36}%
\makeatletter
\providecommand \@ifxundefined [1]{%
 \@ifx{#1\undefined}
}%
\providecommand \@ifnum [1]{%
 \ifnum #1\expandafter \@firstoftwo
 \else \expandafter \@secondoftwo
 \fi
}%
\providecommand \@ifx [1]{%
 \ifx #1\expandafter \@firstoftwo
 \else \expandafter \@secondoftwo
 \fi
}%
\providecommand \natexlab [1]{#1}%
\providecommand \enquote  [1]{``#1''}%
\providecommand \bibnamefont  [1]{#1}%
\providecommand \bibfnamefont [1]{#1}%
\providecommand \citenamefont [1]{#1}%
\providecommand \href@noop [0]{\@secondoftwo}%
\providecommand \href [0]{\begingroup \@sanitize@url \@href}%
\providecommand \@href[1]{\@@startlink{#1}\@@href}%
\providecommand \@@href[1]{\endgroup#1\@@endlink}%
\providecommand \@sanitize@url [0]{\catcode `\\12\catcode `\$12\catcode
  `\&12\catcode `\#12\catcode `\^12\catcode `\_12\catcode `\%12\relax}%
\providecommand \@@startlink[1]{}%
\providecommand \@@endlink[0]{}%
\providecommand \url  [0]{\begingroup\@sanitize@url \@url }%
\providecommand \@url [1]{\endgroup\@href {#1}{\urlprefix }}%
\providecommand \urlprefix  [0]{URL }%
\providecommand \Eprint [0]{\href }%
\providecommand \doibase [0]{http://dx.doi.org/}%
\providecommand \selectlanguage [0]{\@gobble}%
\providecommand \bibinfo  [0]{\@secondoftwo}%
\providecommand \bibfield  [0]{\@secondoftwo}%
\providecommand \translation [1]{[#1]}%
\providecommand \BibitemOpen [0]{}%
\providecommand \bibitemStop [0]{}%
\providecommand \bibitemNoStop [0]{.\EOS\space}%
\providecommand \EOS [0]{\spacefactor3000\relax}%
\providecommand \BibitemShut  [1]{\csname bibitem#1\endcsname}%
\let\auto@bib@innerbib\@empty
\bibitem [{\citenamefont {Fukushima}\ and\ \citenamefont
  {Hatsuda}(2011)}]{Fukushima:2010bq}%
  \BibitemOpen
  \bibfield  {author} {\bibinfo {author} {\bibfnamefont {K.}~\bibnamefont
  {Fukushima}}\ and\ \bibinfo {author} {\bibfnamefont {T.}~\bibnamefont
  {Hatsuda}},\ }\href {\doibase 10.1088/0034-4885/74/1/014001} {\bibfield
  {journal} {\bibinfo  {journal} {Rept. Prog. Phys.}\ }\textbf {\bibinfo
  {volume} {74}},\ \bibinfo {pages} {014001} (\bibinfo {year} {2011})},\
  \Eprint {http://arxiv.org/abs/1005.4814} {arXiv:1005.4814 [hep-ph]}
  \BibitemShut {NoStop}%
\bibitem [{\citenamefont {Lee}\ and\ \citenamefont {Yang}(1952)}]{Lee:1952ig}%
  \BibitemOpen
  \bibfield  {author} {\bibinfo {author} {\bibfnamefont {T.~D.}\ \bibnamefont
  {Lee}}\ and\ \bibinfo {author} {\bibfnamefont {C.-N.}\ \bibnamefont {Yang}},\
  }\href {\doibase 10.1103/PhysRev.87.410} {\bibfield  {journal} {\bibinfo
  {journal} {Phys. Rev.}\ }\textbf {\bibinfo {volume} {87}},\ \bibinfo {pages}
  {410} (\bibinfo {year} {1952})}\BibitemShut {NoStop}%
\bibitem [{\citenamefont {Biskup}\ \emph {et~al.}(2000)\citenamefont {Biskup},
  \citenamefont {Borgs}, \citenamefont {Chayes}, \citenamefont {Kleinwaks},\
  and\ \citenamefont {Koteck{\`y}}}]{biskup2000general}%
  \BibitemOpen
  \bibfield  {author} {\bibinfo {author} {\bibfnamefont {M.}~\bibnamefont
  {Biskup}}, \bibinfo {author} {\bibfnamefont {C.}~\bibnamefont {Borgs}},
  \bibinfo {author} {\bibfnamefont {J.~T.}\ \bibnamefont {Chayes}}, \bibinfo
  {author} {\bibfnamefont {L.~J.}\ \bibnamefont {Kleinwaks}}, \ and\ \bibinfo
  {author} {\bibfnamefont {R.}~\bibnamefont {Koteck{\`y}}},\ }\href@noop {}
  {\bibfield  {journal} {\bibinfo  {journal} {Physical Review Letters}\
  }\textbf {\bibinfo {volume} {84}},\ \bibinfo {pages} {4794} (\bibinfo {year}
  {2000})}\BibitemShut {NoStop}%
\bibitem [{\citenamefont {Peng}\ \emph {et~al.}(2015)\citenamefont {Peng},
  \citenamefont {Zhou}, \citenamefont {Wei}, \citenamefont {Cui}, \citenamefont
  {Du},\ and\ \citenamefont {Liu}}]{peng2015experimental}%
  \BibitemOpen
  \bibfield  {author} {\bibinfo {author} {\bibfnamefont {X.}~\bibnamefont
  {Peng}}, \bibinfo {author} {\bibfnamefont {H.}~\bibnamefont {Zhou}}, \bibinfo
  {author} {\bibfnamefont {B.-B.}\ \bibnamefont {Wei}}, \bibinfo {author}
  {\bibfnamefont {J.}~\bibnamefont {Cui}}, \bibinfo {author} {\bibfnamefont
  {J.}~\bibnamefont {Du}}, \ and\ \bibinfo {author} {\bibfnamefont {R.-B.}\
  \bibnamefont {Liu}},\ }\href@noop {} {\bibfield  {journal} {\bibinfo
  {journal} {Physical review letters}\ }\textbf {\bibinfo {volume} {114}},\
  \bibinfo {pages} {010601} (\bibinfo {year} {2015})}\BibitemShut {NoStop}%
\bibitem [{\citenamefont {Krishnan}\ \emph {et~al.}(2019)\citenamefont
  {Krishnan}, \citenamefont {Schmitt}, \citenamefont {Moessner},\ and\
  \citenamefont {Heyl}}]{PhysRevA.100.022125}%
  \BibitemOpen
  \bibfield  {author} {\bibinfo {author} {\bibfnamefont {A.}~\bibnamefont
  {Krishnan}}, \bibinfo {author} {\bibfnamefont {M.}~\bibnamefont {Schmitt}},
  \bibinfo {author} {\bibfnamefont {R.}~\bibnamefont {Moessner}}, \ and\
  \bibinfo {author} {\bibfnamefont {M.}~\bibnamefont {Heyl}},\ }\href {\doibase
  10.1103/PhysRevA.100.022125} {\bibfield  {journal} {\bibinfo  {journal}
  {Phys. Rev. A}\ }\textbf {\bibinfo {volume} {100}},\ \bibinfo {pages}
  {022125} (\bibinfo {year} {2019})},\ \Eprint
  {http://arxiv.org/abs/1902.07155} {arXiv:1902.07155 [quant-ph]} \BibitemShut
  {NoStop}%
\bibitem [{\citenamefont {Nakamura}\ and\ \citenamefont
  {Nagata}(2016)}]{Nakamura:2013ska}%
  \BibitemOpen
  \bibfield  {author} {\bibinfo {author} {\bibfnamefont {A.}~\bibnamefont
  {Nakamura}}\ and\ \bibinfo {author} {\bibfnamefont {K.}~\bibnamefont
  {Nagata}},\ }\href {\doibase 10.1093/ptep/ptw013} {\bibfield  {journal}
  {\bibinfo  {journal} {PTEP}\ }\textbf {\bibinfo {volume} {2016}},\ \bibinfo
  {pages} {033D01} (\bibinfo {year} {2016})},\ \Eprint
  {http://arxiv.org/abs/1305.0760} {arXiv:1305.0760 [hep-ph]} \BibitemShut
  {NoStop}%
\bibitem [{\citenamefont {Nagata}\ \emph {et~al.}(2015)\citenamefont {Nagata},
  \citenamefont {Kashiwa}, \citenamefont {Nakamura},\ and\ \citenamefont
  {Nishigaki}}]{Nagata:2014fra}%
  \BibitemOpen
  \bibfield  {author} {\bibinfo {author} {\bibfnamefont {K.}~\bibnamefont
  {Nagata}}, \bibinfo {author} {\bibfnamefont {K.}~\bibnamefont {Kashiwa}},
  \bibinfo {author} {\bibfnamefont {A.}~\bibnamefont {Nakamura}}, \ and\
  \bibinfo {author} {\bibfnamefont {S.~M.}\ \bibnamefont {Nishigaki}},\ }\href
  {\doibase 10.1103/PhysRevD.91.094507} {\bibfield  {journal} {\bibinfo
  {journal} {Phys. Rev.}\ }\textbf {\bibinfo {volume} {D91}},\ \bibinfo {pages}
  {094507} (\bibinfo {year} {2015})},\ \Eprint {http://arxiv.org/abs/1410.0783}
  {arXiv:1410.0783 [hep-lat]} \BibitemShut {NoStop}%
\bibitem [{\citenamefont {Wakayama}\ \emph {et~al.}(2019)\citenamefont
  {Wakayama}, \citenamefont {Bornyakov}, \citenamefont {Boyda}, \citenamefont
  {Goy}, \citenamefont {Iida}, \citenamefont {Molochkov}, \citenamefont
  {Nakamura},\ and\ \citenamefont {Zakharov}}]{Wakayama:2018wkc}%
  \BibitemOpen
  \bibfield  {author} {\bibinfo {author} {\bibfnamefont {M.}~\bibnamefont
  {Wakayama}}, \bibinfo {author} {\bibfnamefont {V.~G.}\ \bibnamefont
  {Bornyakov}}, \bibinfo {author} {\bibfnamefont {D.~L.}\ \bibnamefont
  {Boyda}}, \bibinfo {author} {\bibfnamefont {V.~A.}\ \bibnamefont {Goy}},
  \bibinfo {author} {\bibfnamefont {H.}~\bibnamefont {Iida}}, \bibinfo {author}
  {\bibfnamefont {A.~V.}\ \bibnamefont {Molochkov}}, \bibinfo {author}
  {\bibfnamefont {A.}~\bibnamefont {Nakamura}}, \ and\ \bibinfo {author}
  {\bibfnamefont {V.~I.}\ \bibnamefont {Zakharov}},\ }\href {\doibase
  10.1016/j.physletb.2019.04.040} {\bibfield  {journal} {\bibinfo  {journal}
  {Phys. Lett.}\ }\textbf {\bibinfo {volume} {B793}},\ \bibinfo {pages} {227}
  (\bibinfo {year} {2019})},\ \Eprint {http://arxiv.org/abs/1802.02014}
  {arXiv:1802.02014 [hep-lat]} \BibitemShut {NoStop}%
\bibitem [{\citenamefont {Roberge}\ and\ \citenamefont
  {Weiss}(1986)}]{Roberge:1986mm}%
  \BibitemOpen
  \bibfield  {author} {\bibinfo {author} {\bibfnamefont {A.}~\bibnamefont
  {Roberge}}\ and\ \bibinfo {author} {\bibfnamefont {N.}~\bibnamefont
  {Weiss}},\ }\href {\doibase 10.1016/0550-3213(86)90582-1} {\bibfield
  {journal} {\bibinfo  {journal} {Nucl.Phys.}\ }\textbf {\bibinfo {volume}
  {B275}},\ \bibinfo {pages} {734} (\bibinfo {year} {1986})}\BibitemShut
  {NoStop}%
\bibitem [{\citenamefont {Fukuda}\ \emph {et~al.}(2016)\citenamefont {Fukuda},
  \citenamefont {Nakamura},\ and\ \citenamefont {Oka}}]{Fukuda:2015mva}%
  \BibitemOpen
  \bibfield  {author} {\bibinfo {author} {\bibfnamefont {R.}~\bibnamefont
  {Fukuda}}, \bibinfo {author} {\bibfnamefont {A.}~\bibnamefont {Nakamura}}, \
  and\ \bibinfo {author} {\bibfnamefont {S.}~\bibnamefont {Oka}},\ }\href
  {\doibase 10.1103/PhysRevD.93.094508} {\bibfield  {journal} {\bibinfo
  {journal} {Phys. Rev.}\ }\textbf {\bibinfo {volume} {D93}},\ \bibinfo {pages}
  {094508} (\bibinfo {year} {2016})},\ \Eprint
  {http://arxiv.org/abs/1504.06351} {arXiv:1504.06351 [hep-lat]} \BibitemShut
  {NoStop}%
\bibitem [{\citenamefont {de~Forcrand}(2009)}]{deForcrand:2010ys}%
  \BibitemOpen
  \bibfield  {author} {\bibinfo {author} {\bibfnamefont {P.}~\bibnamefont
  {de~Forcrand}},\ }\href@noop {} {\bibfield  {journal} {\bibinfo  {journal}
  {PoS}\ }\textbf {\bibinfo {volume} {LAT2009}},\ \bibinfo {pages} {010}
  (\bibinfo {year} {2009})},\ \Eprint {http://arxiv.org/abs/1005.0539}
  {arXiv:1005.0539 [hep-lat]} \BibitemShut {NoStop}%
\bibitem [{\citenamefont {Kashiwa}(2019)}]{Kashiwa:2019ihm}%
  \BibitemOpen
  \bibfield  {author} {\bibinfo {author} {\bibfnamefont {K.}~\bibnamefont
  {Kashiwa}},\ }\href {\doibase 10.3390/sym11040562} {\bibfield  {journal}
  {\bibinfo  {journal} {Symmetry}\ }\textbf {\bibinfo {volume} {11}},\ \bibinfo
  {pages} {562} (\bibinfo {year} {2019})}\BibitemShut {NoStop}%
\bibitem [{\citenamefont {Kashiwa}\ and\ \citenamefont
  {Ohnishi}(2015)}]{Kashiwa:2015tna}%
  \BibitemOpen
  \bibfield  {author} {\bibinfo {author} {\bibfnamefont {K.}~\bibnamefont
  {Kashiwa}}\ and\ \bibinfo {author} {\bibfnamefont {A.}~\bibnamefont
  {Ohnishi}},\ }\href {\doibase 10.1016/j.physletb.2015.09.036} {\bibfield
  {journal} {\bibinfo  {journal} {Phys. Lett.}\ }\textbf {\bibinfo {volume}
  {B750}},\ \bibinfo {pages} {282} (\bibinfo {year} {2015})},\ \Eprint
  {http://arxiv.org/abs/1505.06799} {arXiv:1505.06799 [hep-ph]} \BibitemShut
  {NoStop}%
\bibitem [{\citenamefont {Kashiwa}\ and\ \citenamefont
  {Ohnishi}(2017)}]{Kashiwa:2017yvy}%
  \BibitemOpen
  \bibfield  {author} {\bibinfo {author} {\bibfnamefont {K.}~\bibnamefont
  {Kashiwa}}\ and\ \bibinfo {author} {\bibfnamefont {A.}~\bibnamefont
  {Ohnishi}},\ }\href {\doibase 10.1016/j.physletb.2017.07.033} {\bibfield
  {journal} {\bibinfo  {journal} {Phys. Lett.}\ }\textbf {\bibinfo {volume}
  {B772}},\ \bibinfo {pages} {669} (\bibinfo {year} {2017})},\ \Eprint
  {http://arxiv.org/abs/1701.04953} {arXiv:1701.04953 [hep-ph]} \BibitemShut
  {NoStop}%
\bibitem [{\citenamefont {Mori}\ \emph {et~al.}(2018)\citenamefont {Mori},
  \citenamefont {Kashiwa},\ and\ \citenamefont {Ohnishi}}]{Mori:2017zyl}%
  \BibitemOpen
  \bibfield  {author} {\bibinfo {author} {\bibfnamefont {Y.}~\bibnamefont
  {Mori}}, \bibinfo {author} {\bibfnamefont {K.}~\bibnamefont {Kashiwa}}, \
  and\ \bibinfo {author} {\bibfnamefont {A.}~\bibnamefont {Ohnishi}},\ }\href
  {\doibase 10.1016/j.physletb.2018.04.018} {\bibfield  {journal} {\bibinfo
  {journal} {Phys. Lett. B}\ }\textbf {\bibinfo {volume} {781}},\ \bibinfo
  {pages} {688} (\bibinfo {year} {2018})},\ \Eprint
  {http://arxiv.org/abs/1705.03646} {arXiv:1705.03646 [hep-lat]} \BibitemShut
  {NoStop}%
\bibitem [{\citenamefont {Mori}\ \emph {et~al.}(2017)\citenamefont {Mori},
  \citenamefont {Kashiwa},\ and\ \citenamefont {Ohnishi}}]{Mori:2017pne}%
  \BibitemOpen
  \bibfield  {author} {\bibinfo {author} {\bibfnamefont {Y.}~\bibnamefont
  {Mori}}, \bibinfo {author} {\bibfnamefont {K.}~\bibnamefont {Kashiwa}}, \
  and\ \bibinfo {author} {\bibfnamefont {A.}~\bibnamefont {Ohnishi}},\ }\href
  {\doibase 10.1103/PhysRevD.96.111501} {\bibfield  {journal} {\bibinfo
  {journal} {Phys. Rev. D}\ }\textbf {\bibinfo {volume} {96}},\ \bibinfo
  {pages} {111501} (\bibinfo {year} {2017})},\ \Eprint
  {http://arxiv.org/abs/1705.05605} {arXiv:1705.05605 [hep-lat]} \BibitemShut
  {NoStop}%
\bibitem [{\citenamefont {Kashiwa}\ \emph
  {et~al.}(2019{\natexlab{a}})\citenamefont {Kashiwa}, \citenamefont {Mori},\
  and\ \citenamefont {Ohnishi}}]{Kashiwa:2018vxr}%
  \BibitemOpen
  \bibfield  {author} {\bibinfo {author} {\bibfnamefont {K.}~\bibnamefont
  {Kashiwa}}, \bibinfo {author} {\bibfnamefont {Y.}~\bibnamefont {Mori}}, \
  and\ \bibinfo {author} {\bibfnamefont {A.}~\bibnamefont {Ohnishi}},\ }\href
  {\doibase 10.1103/PhysRevD.99.014033} {\bibfield  {journal} {\bibinfo
  {journal} {Phys. Rev.}\ }\textbf {\bibinfo {volume} {D99}},\ \bibinfo {pages}
  {014033} (\bibinfo {year} {2019}{\natexlab{a}})},\ \Eprint
  {http://arxiv.org/abs/1805.08940} {arXiv:1805.08940 [hep-ph]} \BibitemShut
  {NoStop}%
\bibitem [{\citenamefont {Kashiwa}\ \emph
  {et~al.}(2019{\natexlab{b}})\citenamefont {Kashiwa}, \citenamefont {Mori},\
  and\ \citenamefont {Ohnishi}}]{Kashiwa:2019lkv}%
  \BibitemOpen
  \bibfield  {author} {\bibinfo {author} {\bibfnamefont {K.}~\bibnamefont
  {Kashiwa}}, \bibinfo {author} {\bibfnamefont {Y.}~\bibnamefont {Mori}}, \
  and\ \bibinfo {author} {\bibfnamefont {A.}~\bibnamefont {Ohnishi}},\ }\href
  {\doibase 10.1103/PhysRevD.99.114005} {\bibfield  {journal} {\bibinfo
  {journal} {Phys. Rev.}\ }\textbf {\bibinfo {volume} {D99}},\ \bibinfo {pages}
  {114005} (\bibinfo {year} {2019}{\natexlab{b}})},\ \Eprint
  {http://arxiv.org/abs/1903.03679} {arXiv:1903.03679 [hep-lat]} \BibitemShut
  {NoStop}%
\bibitem [{\citenamefont {Mori}\ \emph {et~al.}(2019)\citenamefont {Mori},
  \citenamefont {Kashiwa},\ and\ \citenamefont {Ohnishi}}]{Mori:2019tux}%
  \BibitemOpen
  \bibfield  {author} {\bibinfo {author} {\bibfnamefont {Y.}~\bibnamefont
  {Mori}}, \bibinfo {author} {\bibfnamefont {K.}~\bibnamefont {Kashiwa}}, \
  and\ \bibinfo {author} {\bibfnamefont {A.}~\bibnamefont {Ohnishi}},\ }\href
  {\doibase 10.1093/ptep/ptz111} {\bibfield  {journal} {\bibinfo  {journal}
  {PTEP}\ }\textbf {\bibinfo {volume} {2019}},\ \bibinfo {pages} {113B01}
  (\bibinfo {year} {2019})},\ \Eprint {http://arxiv.org/abs/1904.11140}
  {arXiv:1904.11140 [hep-lat]} \BibitemShut {NoStop}%
\bibitem [{\citenamefont {Pawlowski}\ \emph {et~al.}(2020)\citenamefont
  {Pawlowski}, \citenamefont {Scherzer}, \citenamefont {Schmidt}, \citenamefont
  {Ziegler},\ and\ \citenamefont {Ziesch^^c3^^a9}}]{Pawlowski:2020kok}%
  \BibitemOpen
  \bibfield  {author} {\bibinfo {author} {\bibfnamefont {J.~M.}\ \bibnamefont
  {Pawlowski}}, \bibinfo {author} {\bibfnamefont {M.}~\bibnamefont {Scherzer}},
  \bibinfo {author} {\bibfnamefont {C.}~\bibnamefont {Schmidt}}, \bibinfo
  {author} {\bibfnamefont {F.~P.~G.}\ \bibnamefont {Ziegler}}, \ and\ \bibinfo
  {author} {\bibfnamefont {F.}~\bibnamefont {Ziesch^^c3^^a9}},\ }in\ \href@noop
  {} {\emph {\bibinfo {booktitle} {{37th International Symposium on Lattice
  Field Theory (Lattice 2019) Wuhan, Hubei, China, June 16-22, 2019}}}}\
  (\bibinfo {year} {2020})\ \Eprint {http://arxiv.org/abs/2001.09767}
  {arXiv:2001.09767 [hep-lat]} \BibitemShut {NoStop}%
\bibitem [{\citenamefont {Detmold}\ \emph {et~al.}(2020)\citenamefont
  {Detmold}, \citenamefont {Kanwar}, \citenamefont {Wagman},\ and\
  \citenamefont {Warrington}}]{Detmold:2020ncp}%
  \BibitemOpen
  \bibfield  {author} {\bibinfo {author} {\bibfnamefont {W.}~\bibnamefont
  {Detmold}}, \bibinfo {author} {\bibfnamefont {G.}~\bibnamefont {Kanwar}},
  \bibinfo {author} {\bibfnamefont {M.~L.}\ \bibnamefont {Wagman}}, \ and\
  \bibinfo {author} {\bibfnamefont {N.~C.}\ \bibnamefont {Warrington}},\ }\href
  {\doibase 10.1103/PhysRevD.102.014514} {\bibfield  {journal} {\bibinfo
  {journal} {Phys. Rev. D}\ }\textbf {\bibinfo {volume} {102}},\ \bibinfo
  {pages} {014514} (\bibinfo {year} {2020})}\BibitemShut {NoStop}%
\bibitem [{\citenamefont {Wilson}(1974)}]{Wilson:1974sk}%
  \BibitemOpen
  \bibfield  {author} {\bibinfo {author} {\bibfnamefont {K.~G.}\ \bibnamefont
  {Wilson}},\ }\href {\doibase 10.1103/PhysRevD.10.2445} {\bibfield  {journal}
  {\bibinfo  {journal} {Phys. Rev.}\ }\textbf {\bibinfo {volume} {D10}},\
  \bibinfo {pages} {2445} (\bibinfo {year} {1974})},\ \bibinfo {note}
  {[,45(1974); ,319(1974)]}\BibitemShut {NoStop}%
\bibitem [{\citenamefont {Balian}\ \emph {et~al.}(1974)\citenamefont {Balian},
  \citenamefont {Drouffe},\ and\ \citenamefont {Itzykson}}]{Balian:1974ts}%
  \BibitemOpen
  \bibfield  {author} {\bibinfo {author} {\bibfnamefont {R.}~\bibnamefont
  {Balian}}, \bibinfo {author} {\bibfnamefont {J.~M.}\ \bibnamefont {Drouffe}},
  \ and\ \bibinfo {author} {\bibfnamefont {C.}~\bibnamefont {Itzykson}},\
  }\href {\doibase 10.1103/PhysRevD.10.3376} {\bibfield  {journal} {\bibinfo
  {journal} {Phys. Rev.}\ }\textbf {\bibinfo {volume} {D10}},\ \bibinfo {pages}
  {3376} (\bibinfo {year} {1974})}\BibitemShut {NoStop}%
\bibitem [{\citenamefont {Alexandru}\ \emph {et~al.}(2016)\citenamefont
  {Alexandru}, \citenamefont {Basar}, \citenamefont {Bedaque}, \citenamefont
  {Ridgway},\ and\ \citenamefont {Warrington}}]{Alexandru:2015sua}%
  \BibitemOpen
  \bibfield  {author} {\bibinfo {author} {\bibfnamefont {A.}~\bibnamefont
  {Alexandru}}, \bibinfo {author} {\bibfnamefont {G.}~\bibnamefont {Basar}},
  \bibinfo {author} {\bibfnamefont {P.~F.}\ \bibnamefont {Bedaque}}, \bibinfo
  {author} {\bibfnamefont {G.~W.}\ \bibnamefont {Ridgway}}, \ and\ \bibinfo
  {author} {\bibfnamefont {N.~C.}\ \bibnamefont {Warrington}},\ }\href
  {\doibase 10.1007/JHEP05(2016)053} {\bibfield  {journal} {\bibinfo  {journal}
  {JHEP}\ }\textbf {\bibinfo {volume} {05}},\ \bibinfo {pages} {053} (\bibinfo
  {year} {2016})},\ \Eprint {http://arxiv.org/abs/1512.08764} {arXiv:1512.08764
  [hep-lat]} \BibitemShut {NoStop}%
\bibitem [{\citenamefont {Alexandru}\ \emph {et~al.}(2017)\citenamefont
  {Alexandru}, \citenamefont {Bedaque}, \citenamefont {Lamm},\ and\
  \citenamefont {Lawrence}}]{Alexandru:2017czx}%
  \BibitemOpen
  \bibfield  {author} {\bibinfo {author} {\bibfnamefont {A.}~\bibnamefont
  {Alexandru}}, \bibinfo {author} {\bibfnamefont {P.~F.}\ \bibnamefont
  {Bedaque}}, \bibinfo {author} {\bibfnamefont {H.}~\bibnamefont {Lamm}}, \
  and\ \bibinfo {author} {\bibfnamefont {S.}~\bibnamefont {Lawrence}},\ }\href
  {\doibase 10.1103/PhysRevD.96.094505} {\bibfield  {journal} {\bibinfo
  {journal} {Phys. Rev.}\ }\textbf {\bibinfo {volume} {D96}},\ \bibinfo {pages}
  {094505} (\bibinfo {year} {2017})},\ \Eprint
  {http://arxiv.org/abs/1709.01971} {arXiv:1709.01971 [hep-lat]} \BibitemShut
  {NoStop}%
\bibitem [{\citenamefont {Zeiler}(2012)}]{zeiler2012adadelta}%
  \BibitemOpen
  \bibfield  {author} {\bibinfo {author} {\bibfnamefont {M.~D.}\ \bibnamefont
  {Zeiler}},\ }\href@noop {} {\bibfield  {journal} {\bibinfo  {journal} {arXiv
  preprint arXiv:1212.5701}\ } (\bibinfo {year} {2012})}\BibitemShut {NoStop}%
\bibitem [{\citenamefont {Glorot}\ and\ \citenamefont
  {Bengio}(2010)}]{glorot2010understanding}%
  \BibitemOpen
  \bibfield  {author} {\bibinfo {author} {\bibfnamefont {X.}~\bibnamefont
  {Glorot}}\ and\ \bibinfo {author} {\bibfnamefont {Y.}~\bibnamefont
  {Bengio}},\ }in\ \href@noop {} {\emph {\bibinfo {booktitle} {Proceedings of
  the Thirteenth International Conference on Artificial Intelligence and
  Statistics}}}\ (\bibinfo {year} {2010})\ pp.\ \bibinfo {pages}
  {249--256}\BibitemShut {NoStop}%
\bibitem [{\citenamefont {Ioffe}\ and\ \citenamefont
  {Szegedy}(2015)}]{ioffe2015batch}%
  \BibitemOpen
  \bibfield  {author} {\bibinfo {author} {\bibfnamefont {S.}~\bibnamefont
  {Ioffe}}\ and\ \bibinfo {author} {\bibfnamefont {C.}~\bibnamefont
  {Szegedy}},\ }\href@noop {} {\  (\bibinfo {year} {2015})},\ \Eprint
  {http://arxiv.org/abs/1502.03167} {arXiv:1502.03167 [cs.LG]} \BibitemShut
  {NoStop}%
\bibitem [{\citenamefont {Duane}\ \emph {et~al.}(1987)\citenamefont {Duane},
  \citenamefont {Kennedy}, \citenamefont {Pendleton},\ and\ \citenamefont
  {Roweth}}]{Duane:1987de}%
  \BibitemOpen
  \bibfield  {author} {\bibinfo {author} {\bibfnamefont {S.}~\bibnamefont
  {Duane}}, \bibinfo {author} {\bibfnamefont {A.~D.}\ \bibnamefont {Kennedy}},
  \bibinfo {author} {\bibfnamefont {B.~J.}\ \bibnamefont {Pendleton}}, \ and\
  \bibinfo {author} {\bibfnamefont {D.}~\bibnamefont {Roweth}},\ }\href
  {\doibase 10.1016/0370-2693(87)91197-X} {\bibfield  {journal} {\bibinfo
  {journal} {Phys. Lett.}\ }\textbf {\bibinfo {volume} {B195}},\ \bibinfo
  {pages} {216} (\bibinfo {year} {1987})}\BibitemShut {NoStop}%
\bibitem [{\citenamefont {Swendsen}\ and\ \citenamefont
  {Wang}(1986)}]{swendsen1986replica}%
  \BibitemOpen
  \bibfield  {author} {\bibinfo {author} {\bibfnamefont {R.~H.}\ \bibnamefont
  {Swendsen}}\ and\ \bibinfo {author} {\bibfnamefont {J.-S.}\ \bibnamefont
  {Wang}},\ }\href@noop {} {\bibfield  {journal} {\bibinfo  {journal} {Physical
  review letters}\ }\textbf {\bibinfo {volume} {57}},\ \bibinfo {pages} {2607}
  (\bibinfo {year} {1986})}\BibitemShut {NoStop}%
\bibitem [{\citenamefont {Fukuma}\ and\ \citenamefont
  {Umeda}(2017)}]{Fukuma:2017fjq}%
  \BibitemOpen
  \bibfield  {author} {\bibinfo {author} {\bibfnamefont {M.}~\bibnamefont
  {Fukuma}}\ and\ \bibinfo {author} {\bibfnamefont {N.}~\bibnamefont {Umeda}},\
  }\href {\doibase 10.1093/ptep/ptx081} {\bibfield  {journal} {\bibinfo
  {journal} {PTEP}\ }\textbf {\bibinfo {volume} {2017}},\ \bibinfo {pages}
  {073B01} (\bibinfo {year} {2017})},\ \Eprint
  {http://arxiv.org/abs/1703.00861} {arXiv:1703.00861 [hep-lat]} \BibitemShut
  {NoStop}%
\bibitem [{\citenamefont {Fujii}\ \emph {et~al.}(2013)\citenamefont {Fujii},
  \citenamefont {Honda}, \citenamefont {Kato}, \citenamefont {Kikukawa},
  \citenamefont {Komatsu},\ and\ \citenamefont {Sano}}]{Fujii:2013sra}%
  \BibitemOpen
  \bibfield  {author} {\bibinfo {author} {\bibfnamefont {H.}~\bibnamefont
  {Fujii}}, \bibinfo {author} {\bibfnamefont {D.}~\bibnamefont {Honda}},
  \bibinfo {author} {\bibfnamefont {M.}~\bibnamefont {Kato}}, \bibinfo {author}
  {\bibfnamefont {Y.}~\bibnamefont {Kikukawa}}, \bibinfo {author}
  {\bibfnamefont {S.}~\bibnamefont {Komatsu}}, \ and\ \bibinfo {author}
  {\bibfnamefont {T.}~\bibnamefont {Sano}},\ }\href {\doibase
  10.1007/JHEP10(2013)147} {\bibfield  {journal} {\bibinfo  {journal} {JHEP}\
  }\textbf {\bibinfo {volume} {1310}},\ \bibinfo {pages} {147} (\bibinfo {year}
  {2013})},\ \Eprint {http://arxiv.org/abs/1309.4371} {arXiv:1309.4371
  [hep-lat]} \BibitemShut {NoStop}%
\bibitem [{\citenamefont {Alexandru}\ \emph {et~al.}(2018)\citenamefont
  {Alexandru}, \citenamefont {Bedaque}, \citenamefont {Lamm},\ and\
  \citenamefont {Lawrence}}]{Alexandru:2018fqp}%
  \BibitemOpen
  \bibfield  {author} {\bibinfo {author} {\bibfnamefont {A.}~\bibnamefont
  {Alexandru}}, \bibinfo {author} {\bibfnamefont {P.~F.}\ \bibnamefont
  {Bedaque}}, \bibinfo {author} {\bibfnamefont {H.}~\bibnamefont {Lamm}}, \
  and\ \bibinfo {author} {\bibfnamefont {S.}~\bibnamefont {Lawrence}},\ }\href
  {\doibase 10.1103/PhysRevD.97.094510} {\bibfield  {journal} {\bibinfo
  {journal} {Phys. Rev.}\ }\textbf {\bibinfo {volume} {D97}},\ \bibinfo {pages}
  {094510} (\bibinfo {year} {2018})},\ \Eprint
  {http://arxiv.org/abs/1804.00697} {arXiv:1804.00697 [hep-lat]} \BibitemShut
  {NoStop}%
\bibitem [{\citenamefont {Bursa}\ and\ \citenamefont
  {Kroyter}(2018)}]{Bursa:2018ykf}%
  \BibitemOpen
  \bibfield  {author} {\bibinfo {author} {\bibfnamefont {F.}~\bibnamefont
  {Bursa}}\ and\ \bibinfo {author} {\bibfnamefont {M.}~\bibnamefont
  {Kroyter}},\ }\href {\doibase 10.1007/JHEP12(2018)054} {\bibfield  {journal}
  {\bibinfo  {journal} {JHEP}\ }\textbf {\bibinfo {volume} {12}},\ \bibinfo
  {pages} {054} (\bibinfo {year} {2018})},\ \Eprint
  {http://arxiv.org/abs/1805.04941} {arXiv:1805.04941 [hep-lat]} \BibitemShut
  {NoStop}%
\bibitem [{\citenamefont {Makino}\ \emph {et~al.}(2015)\citenamefont {Makino},
  \citenamefont {Suzuki},\ and\ \citenamefont {Takeda}}]{Makino:2015ooa}%
  \BibitemOpen
  \bibfield  {author} {\bibinfo {author} {\bibfnamefont {H.}~\bibnamefont
  {Makino}}, \bibinfo {author} {\bibfnamefont {H.}~\bibnamefont {Suzuki}}, \
  and\ \bibinfo {author} {\bibfnamefont {D.}~\bibnamefont {Takeda}},\ }\href
  {\doibase 10.1103/PhysRevD.92.085020} {\bibfield  {journal} {\bibinfo
  {journal} {Phys. Rev.}\ }\textbf {\bibinfo {volume} {D92}},\ \bibinfo {pages}
  {085020} (\bibinfo {year} {2015})},\ \Eprint
  {http://arxiv.org/abs/1503.00417} {arXiv:1503.00417 [hep-lat]} \BibitemShut
  {NoStop}%
\bibitem [{\citenamefont {Choe}\ \emph {et~al.}(2002)\citenamefont {Choe},
  \citenamefont {Muroya}, \citenamefont {Nakamura}, \citenamefont {Nonaka},
  \citenamefont {Saito},\ and\ \citenamefont {Shoji}}]{Choe:2002pu}%
  \BibitemOpen
  \bibfield  {author} {\bibinfo {author} {\bibfnamefont {S.}~\bibnamefont
  {Choe}}, \bibinfo {author} {\bibfnamefont {S.}~\bibnamefont {Muroya}},
  \bibinfo {author} {\bibfnamefont {A.}~\bibnamefont {Nakamura}}, \bibinfo
  {author} {\bibfnamefont {C.}~\bibnamefont {Nonaka}}, \bibinfo {author}
  {\bibfnamefont {T.}~\bibnamefont {Saito}}, \ and\ \bibinfo {author}
  {\bibfnamefont {F.}~\bibnamefont {Shoji}},\ }\href {\doibase
  10.1016/S0920-5632(01)01920-X} {\bibfield  {journal} {\bibinfo  {journal}
  {Nucl. Phys. B Proc. Suppl.}\ }\textbf {\bibinfo {volume} {106}},\ \bibinfo
  {pages} {1037} (\bibinfo {year} {2002})}\BibitemShut {NoStop}%
\end{thebibliography}%

\end{document}